\documentclass[fleqn,usenatbib]{mnras}

\usepackage[T1]{fontenc}
\usepackage{ae,aecompl}

\usepackage{graphicx}	
\usepackage{amsmath}	
\usepackage{amssymb}	
\DeclareGraphicsExtensions{.png,.pdf}

\title{The linear bias of radio galaxies at ${z\approx0.3}$ via cosmic microwave background lensing} 

\author[C. Devereux et. al.]{
C. Devereux,\thanks{E-mail: c.devereux@herts.ac.uk}
J. E. Geach \&
M. J. Hardcastle
\\
Centre for Astrophysics Research, School of Physics, Astronomy \& Mathematics, University of Hertfordshire, Hatfield, AL10 9AB
}

\pubyear{2018}

\begin{document}
\label{firstpage}
\pagerange{\pageref{firstpage}--\pageref{lastpage}}
\maketitle

\begin{abstract}
We present a new measurement of the linear bias of radio loud active galactic nuclei (RLAGN) at $z\approx0.3$ and $L_{\rm 1.4GHz}>10^{23}\,{\rm W\,Hz^{-1}}$ selected from the Best \& Heckman (2012) sample, made by cross-correlating the RLAGN surface density with a map of the convergence of the weak lensing field of the cosmic microwave background from {\it Planck}. We detect the cross-power signal at a significance of $3\sigma$ and use the amplitude of the cross-power spectrum to estimate the linear bias of RLAGN, $b=2.5 \pm 0.8$, corresponding to a typical dark matter halo mass of $\log_{10}(M_{\rm h} /h^{-1} M_\odot)=14.0^{+0.3}_{-0.5}$. When RLAGN associated with optically-selected clusters are removed we measure a lower bias corresponding to $\log_{10}(M_{\rm h} /h^{-1} M_\odot)=13.7^{+0.4}_{-1.0}$. These observations support the view that powerful RLAGN typically inhabit rich group and cluster environments.
\end{abstract}




\section{Introduction}
\label{sec:intro}
Active galactic nuclei (AGN) play a fundamental role in galaxy evolution: they return vast amounts of energy into the interstellar and intergalactic medium through feedback, which can quench star formation and curtail cooling flows (see e.g. \cite{2007MNRAS.376.1849H, 2012NJPh...14e5023M}, and reviews by \cite{2012ARA&A..50..455F} and \cite{2014ARA&A..52..589H}). It is now well known that the mass of central supermassive black holes (SMBHs) that power AGN is strongly correlated with the stellar mass of their host \citep{Ferrarese:2000je, 2004ApJ...604L..89H, 2005MNRAS.362...25B} which in turn is correlated with local environment, for the most massive galaxies tend to reside in the most massive halos. The details of the link between stellar mass growth in galaxies, its dependence on local environment, and the role of AGN in regulating galaxy growth is hard to disentangle. A simple question that can provide important clues is in what environments do the most powerful AGN reside at $z=0$?

Radio-loud AGN (RLAGN) have high radio luminosities, $L_{\rm 1.4GHz}>10^{23}\,{\rm W\,Hz^{-1}}$, and reside in massive galaxies, $M_\star\approx10^{11-12} h^{-1} M_{\odot}$ \citep{1989MNRAS.240..129Y, 1991ApJ...367....1H, 2007MNRAS.379..260M}. RLAGN are often hosted by giant elliptical galaxies which preferentially sit within galaxy clusters \citep{2010MNRAS.407.1078D, 2013ApJ...770..136I} although it is not clear whether a high density environment is necessary for the RLAGN to be triggered \citep{2009MNRAS.394...38P}. It has also been shown that RLAGN sit in more clustered environments than optically-selected quasars \citep{2017MNRAS.464.3271M, 2017A&A...600A..97R}. In particular, there is evidence that optically-selected quasars reside in halos of typical mass of poor galaxy groups $10^{12-13} h^{-1} M_{\odot}$ \citep{2005MNRAS.356..415C, 2012PhRvD..86h3006S, 2013ApJ...776L..41G} and that RLAGN reside in halos consistent with rich galaxy groups and clusters ($M_{\rm h}>10^{13} h^{-1} M_{\odot}$) \citep{2004MNRAS.350.1485M, 2009ApJ...696..891H}. 

The two-point angular correlation is most commonly used to determine the clustering of local radio galaxies, which generally indicate typical halo masses of $M_{\rm h}\approx10^{13.5-14}h^{-1} M_{\odot}$ for RLAGN, consistent with the picture described above. 
The recent high redshift work ($\langle z\rangle \sim 1.3$) of \cite{2017MNRAS.464.3271M}, \cite{2017A&A...600A..97R} and \cite{2018MNRAS.474.4133H} estimate similar halo masses to the low redshift work of \cite{2004MNRAS.350.1485M} ($\langle z\rangle \sim 0.1$), implying little evolution in the typical host halos of RLAGN (although see \cite{2014MNRAS.440.1527L}). 
The technique of galaxy weak lensing has also been used to measure the projected mass density of the halo through the lensing shear of background galaxies by \cite{2009MNRAS.393..377M} who measured a halo mass of $M_{\rm h}=10^{13.3} h^{-1} M_{\odot}$ for RLAGN at z<0.3, mainly an Fanaroff-Riley type I (FRI) \citep{1974MNRAS.167P..31F} sample, that are not known to reside in galaxy clusters. In yet another approach, \cite{2013ApJ...770..136I, 2015MNRAS.453.2682I} measured the X-ray luminosity of the intra-cluster medium (ICM) as a way of characterising the environments of RLAGN, estimating masses of order $M_{\rm h}\sim 10^{14} h^{-1} M_{\odot}$. This study identified a weak correlation between environment and AGN type, with evidence that high-excitation radio galaxies (HERGs) avoid richer environments compared to those of low excitation radio galaxies (LERGs). Evidence that the RLAGN preferentially inhabit the most massive halos was presented by \cite{2009ApJ...696..891H}, who again used a clustering analysis of a high redshift (z<0.8) sample, to measure a typical halo mass for RLAGN of $\log_{10}(M_{\rm h} /h^{-1} M_\odot)=13.4^{+0.1}_{-0.2}$ compared to $\log_{10}(M_{\rm h} /h^{-1} M_\odot)=12.8^{+0.2}_{-0.3}$ for X-ray selected AGN at the same redshift (see also \cite{2013MNRAS.430..661M} and \cite{2015MNRAS.446.1874L}). 

Here we approach the question using a relatively new method, cross-correlation of galaxy populations with the CMB weak lensing field to measure the bias of the population \citep{2012PhRvD..86h3006S, 2012ApJ...753L...9B, 2013ApJ...776L..41G, 2014A&A...571A..17P, 2015MNRAS.446.3492D, 2016MNRAS.456..924D}. 
Previously, \cite{2015MNRAS.451..849A} measured RLAGN using CMB lensing from the Atacama Cosmology Telescope (ACT), yielding a halo mass of $\log_{10}(M_{\rm h} /h^{-1} M_\odot)=13.5^{+0.5}_{-1.5}$ with a high redshift sample. In this work we use a clean sample of RLAGN originally selected by \cite{2012MNRAS.421.1569B} and the most recent CMB lensing map from {\it Planck} \citep{PlanckCollaboration:2018wma}. We describe the sample and methodology in Section 2.1 \& 2.2, and the main result in Section 2.3. Section 3 presents a discussion of the result and our conclusions.  Throughout we adopt a {\it Planck} 2018 cosmology with  $\Omega_m = 0.3111$,  $\Omega_b = 0.0490$, $\Omega _\Lambda = 0.6889$,  $\sigma_8= 0.8102$, and $h=H_0 / 100\,{\rm km\,s^{-1}\,Mpc^{-1}} = 0.6766$.

\begin{figure}
	\includegraphics[width=\columnwidth]{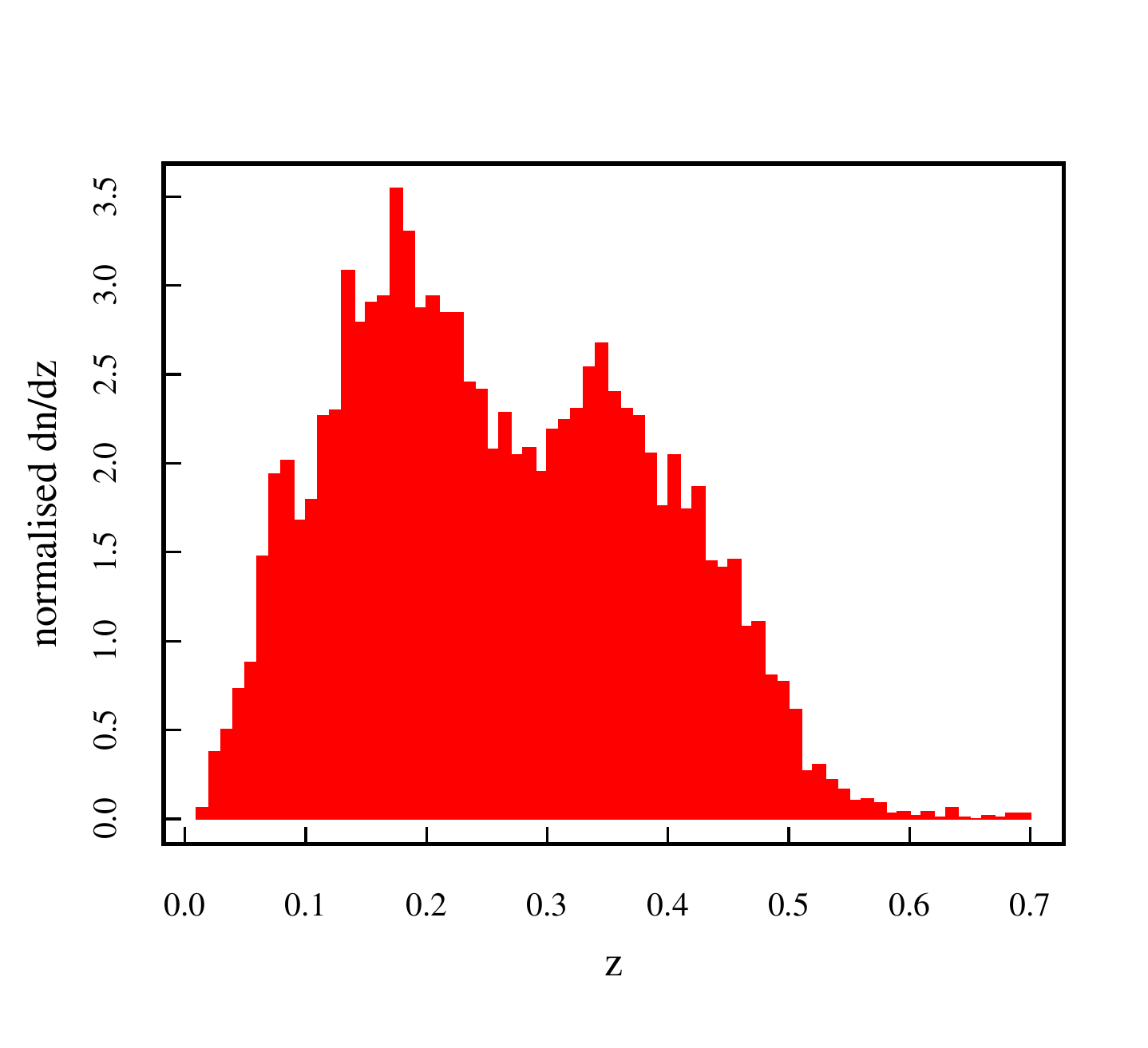}
    \caption{Normalised redshift distribution of the 12,820 RLAGN sample used in this study with a mean redshift of $\langle z\rangle = 0.26$. The sample was taken from the Best and Heckman (2012) selection from the FIRST catalogue and optically identified using SDSS (see Section 2.1).}
    \label{fig:redshift}
\end{figure}

\begin{figure}
	\includegraphics[width=\columnwidth]{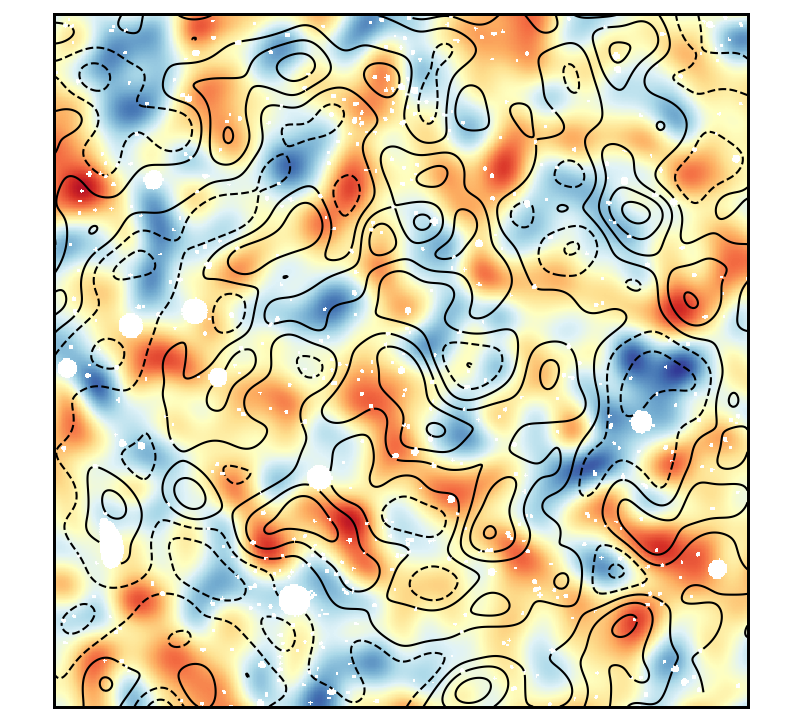}
    \caption{The background image shows the {\it Planck} convergence map spanning 50 degrees centred on $\alpha = 184.6^{\circ}$, $\delta = 32.6^{\circ}$. The contours show the relative RLAGN overdensity where solid contours start at zero and increase in steps of 0.2. Dashed contours are the equivalent levels at negative overdensities. White regions indicate masked areas. For clarity, both maps have been smoothed with a Gaussian kernel of full width at half maximum 3 degrees. Interestingly, the correlation between the maps can be made out by  eye, which we quantify by calculating the cross-power spectrum between the convergence and overdensity maps.}
    \label{fig:overdensitymap}
\end{figure} 

\section{Analysis}
\label{sec:Galaxy data}

\subsection{The Radio-loud AGN sample}
\label{galaxies}

\cite{2012MNRAS.421.1569B} present a catalogue of RLAGN chosen from NVSS \citep{1998AJ....115.1693C} and FIRST \citep{1995ApJ...450..559B} with $S_{\rm 1.4GHz}>5$\,mJy, which has been methodically separated from star forming galaxies. This gives a sample that has less contamination than the full catalogues. Each RLAGN has been optically classified from Sloan Digital Sky Survey (SDSS) spectroscopy (Data Release 7) \citep{2009ApJS..182..543A} and, since matter bias is redshift dependent, having good spectroscopic redshift data makes the result more reliable than estimating the redshift distribution \citep{2015MNRAS.451..849A}.
From the parent sample, we select 12,820 sources classified as AGN, resulting in a sample of RLAGN with $10^{23} \lesssim (L_{\rm 1.4GHz}/{\rm W\,Hz^{-1}}) \lesssim 10^{26.5}$. Figure \ref{fig:redshift} shows the redshift distribution of the sample, which has a mean $\langle z\rangle = 0.26$. We generate a {\sc healpix} map \citep{2005ApJ...622..759G} of the RLAGN surface density through a counts-in-cells method, simply summing the number of sources falling within a particular {\sc healpix} {\tt nside = 2048} pixel and normalising by the solid angle subtended by that pixel, giving the local surface density $\rho$. We then evaluate the fractional overdensity $\delta = {\left(\rho - \langle \rho\rangle\right)}/{\langle\rho\rangle}$, where $\langle \rho\rangle$ is the average density of radio galaxies over the survey. For the latter we determine the total area of the survey mask, which is defined by the union of the SDSS and FIRST survey footprints. This results in a region roughly bounded by Declinations from $-4^\circ$ to $63^\circ$ and Right Ascension $262^\circ$ to $110^\circ$. The total solid angle is $\Omega=6552$\,deg$^2$.

\subsection{CMB weak lensing cross-correlation}
\label{sec:CMB lensing theory}
 
The lensed CMB temperature in direction $\hat{n}$ is related to the unlensed temperature:

\begin{equation}
	T_{\rm lensed}\left(\hat{n}\right) = T_{\rm  unlensed}\left(\hat{n}+\alpha\right)
\end{equation}

\noindent where the deflection angle $\alpha = \nabla\phi\left(\hat{n}\right)$ and $\phi\left(\hat{n}\right)$ is the projected lensing potential. The lensing convergence is $\kappa \approx -\nabla\alpha/2$. In this analysis we use the {\it Planck} 2018 baseline lensing map \citep{PlanckCollaboration:2018wma} which estimates $\kappa$ using a minimum variance quadratic estimator \citep{1999PhRvD..59l3507Z, 2003PhRvD..67h3002O}. An associated mask \citep{2014A&A...571A..17P} removes approximately a third of the sky due to contamination from the Galactic foreground, bright Sunyaev-Zel'dovich Effect clusters and point sources, resulting in a lensing estimate over 67 per cent of the sky. We combine the {\it Planck} lensing mask with the RLAGN survey mask described in section \ref{galaxies} to create a union mask which we use in the following analysis. 

In Figure \ref{fig:overdensitymap} we show a $50^\circ \times 50^\circ$ flat-sky projection of the lensing map with the RLAGN overdensity map overlaid as contours (both maps smoothed with a Gaussian of width $3^\circ$). In the following we quantify the cross-power between the RLAGN density and lensing full (partial sky) maps.

\begin{figure}
	\includegraphics[width=0.5\textwidth]{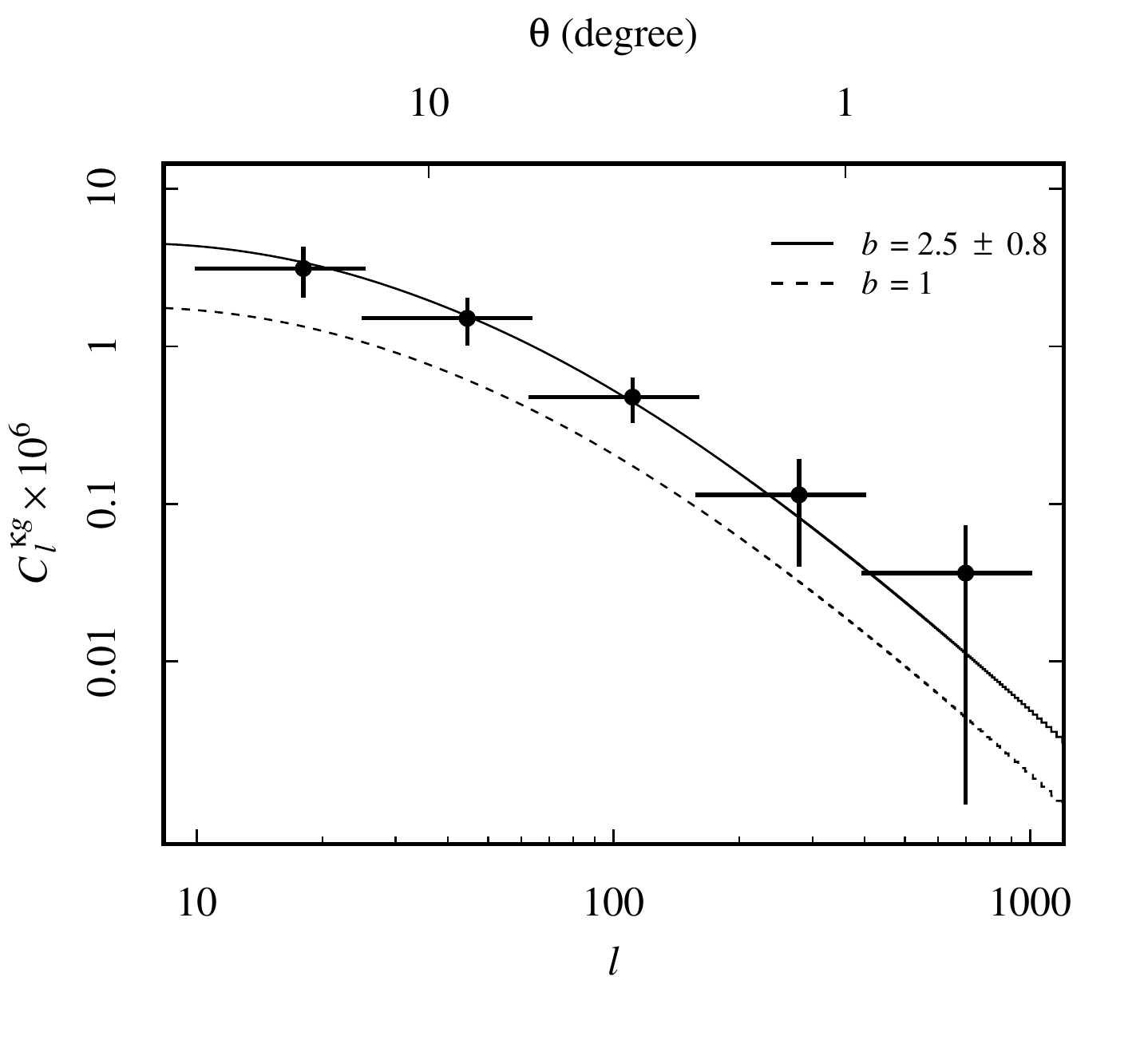}
	\includegraphics[width=0.5\textwidth]{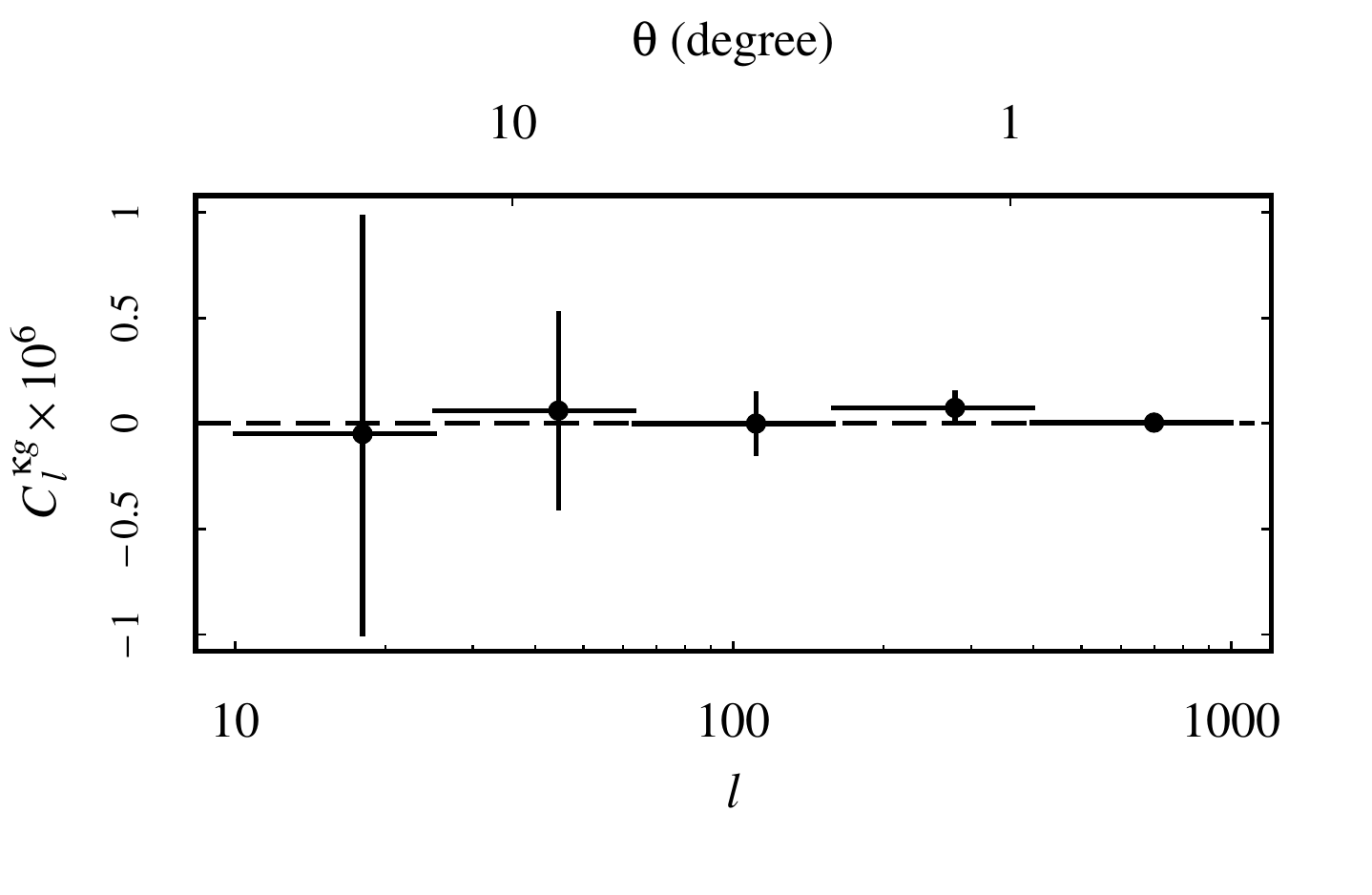}
    \caption{(top) The cross-power spectrum of the overdensities of the Best and Heckman RLAGN and {\it Planck} 2018 CMB lensing. The dashed line shows a model where the linear bias is unity, and the solid line shows the best fit to the data $b = 2.5 \pm 0.8$. The horizontal error bars indicate the bin width in multipole $l$ and the vertical error bars represent the $1\sigma$ uncertainty derived from the scatter in 300 cross-correlations between the RLAGN over-density map and independent $\kappa$ noise realisations. (bottom) The cross-power spectrum measured after the lensing map has been rotated by 90 degrees, note the linear ordinate axis, showing no significant cross-correlation (uncertainties estimated as above).}
    \label{fig:spherebiasplot}
\end{figure}

We calculate the cross-power spectrum using {\sc polspice} \citep{2011ascl.soft09005C}, which employs fast spherical harmonic transforms and allows for a cut-sky approach using a `pseudo-$C_\ell$' estimator technique (\cite{1973ApJ...185..413P}, \cite{2001PhRvD..64h3003W}, \cite{2004MNRAS.349..603E}). The maps are apodized using a cosine weighting function ($\theta_{\rm max}=75^\circ$).  In Figure \ref{fig:spherebiasplot} we present the $\ell$-binned cross-power spectrum.  To estimate the uncertainty on the cross-power spectrum, we use the ensemble of 300 $\kappa$ noise realisations released as part of the {\it Planck} 2018 lensing package. These are based on the unlensed CMB power spectrum combined with artificial lensing potentials to produce maps with projected distributions uncorrelated with the real CMB in the presence of realistic noise (including instrumental noise, Gaussian simulations of foreground power, point-source shot noise). We perform identical cross-correlations between the RLAGN density map and the noise realisations, constructing a covariance matrix:

\begin{equation}
	\mathbf{C}_{i,j}= \frac{1}{(N-1)} \sum_{k=1}^{N=300}(C_{k,i} - \overline{C_i}) (C_{k,j} - \overline{C_j})
    \label{eq:covar}
\end{equation}

\noindent where $i,j$ run over bins in $\ell$ and $\overline{C}$ indicates the average over all $N=300$ noise realisations per bin.

To model the observed cross-power spectrum, we follow the formalism of \cite{2012ApJ...753L...9B}, \cite{2013ApJ...776L..41G} and others, where $C^{\kappa g}_\ell$ is modelled using the Limber approximation \citep{1953ApJ...117..134L, 1992ApJ...388..272K}, which is accurate to about 10\% for scales larger than a few degrees \citep{2007A&A...473..711S}:

\begin{equation}
	C^{\kappa g}_\ell = \int dz \frac{d\eta}{dz} \frac{1}{\eta^2} W^\kappa\left(\eta\right) W^g\left(\eta\right) \mathcal{P}\left(\frac{\ell}{\eta}, z\right)
    \label{eq:Limber}
\end{equation} 

\noindent where $\mathcal{P}(\ell/\eta, z)$ is the linear matter power spectrum \citep{1999ApJ...511....5E}, $W^\kappa\left(\eta\right)$ is the lensing convergence kernel, $W^g\left(\eta\right)$ is the RLAGN distribution kernel and $\eta$ is the comoving distance to redshift $z$. The lensing kernel is given by \citep{2000ApJ...534..533C, 2003ApJ...590..664S}

\begin{equation}
	W^\kappa \left(\eta\right) = \frac{3}{2} \Omega_{\rm m} H_0^2 \frac{\eta}{a(\eta)} \frac{\eta_{\rm CMB} - \eta} {\eta_{\rm CMB}}
    \label{eq:Wk}
\end{equation} 

\noindent  where $a(\eta)$ is the cosmological scale factor, with $\eta_{\rm CMB} \approx 14$\,Gpc. The convergence $\kappa$ along a particular line-of-sight $\hat{n}$ is related to $\delta$ by

\begin{equation}
	\kappa(\hat{n}) = \int d\eta W^\kappa(\eta) \delta(\eta\hat{n}, z)
    \label{eq:convergence}
\end{equation}

\noindent Similarly, fluctuations in the RLAGN density field can be represented by  

\begin{equation}
	g(\hat{n}) = \int d\eta W^g(\eta) \delta(\eta\hat{n}, z)
    \label{eq:AGNdensity}
\end{equation}

\noindent where the AGN distribution kernel is given by

\begin{equation}
	W^g(\eta) = \frac{dz}{d\eta} \frac{dn}{dz} b(\eta)
    \label{eq:Wg}
\end{equation}

\noindent and ${dn}/{dz}$ is the integral-normalized redshift distribution of the  sample. Finally, $b$ is the linear bias of the galaxies.

\subsection{Results}

The bias $b$ sets the amplitude of the cross-power spectrum, and we can estimate it by minimizing 

\begin{equation}
	\chi^2 = (C_\ell^{\rm obs} - C_\ell^{\rm model}) ^ {\rm T} \mathbf{C}^{-1}(C_\ell^{\rm obs} - C_\ell^{\rm model}).
\end{equation}

\noindent We measure $b=2.5\pm 0.8$ with $\chi^2=0.3$. For the null hypothesis $b=0$, the significance of the detection is given by $\Delta\chi^2 = \chi^2_{\rm null} - \chi^2$. We measure $\chi^2_{\rm null}=10.2$, indicating a detection significance of 3.2$\sigma$ for the cross-power signal. The cross-power spectrum for $b=2.5$ is shown in Figure \ref{fig:spherebiasplot}. We perform an additional check by rotating the CMB lensing map by $90^\circ$ to misalign the maps before measuring the RLAGN-lensing cross-correlation, confirming a null detection (see Figure \ref{fig:spherebiasplot}).

We use the bias to estimate the characteristic or average halo mass, $M_{\rm h}$, using the bias-halo fitting function of \cite{2010ApJ...724..878T}, and assuming the halo mass is defined as the total mass enclosed with a radius within which the average density is 200 times the mean density of the Universe. The function is defined in terms of the ratio of the critical density for spherical collapse $\delta_c$ and the variance of the matter field on scales of the halo, $\sigma(R)$, where $R=(3M_{\rm h}/4\pi\rho_m)^{1/3}$ and $\rho_m$ is the mean density of the Universe. Using our measured bias, we find a characteristic RLAGN halo mass of $\log_{10}(M_{\rm h} /h^{-1} M_\odot)=14.0^{+0.3}_{-0.5}$ at $z = 0.26$.

\section{Discussion \& Conclusions}
\label{sec:results}

The initial implications of these results is that RLAGN reside in rich galaxy groups and clusters. This supports the hypothesis \citep{2017MNRAS.464.3271M} that RLAGN require massive galaxies, since the most massive galaxies will preferentially sit in group/cluster-scale halos. How does our measurement compare to the literature? Using the ACT CMB lensing map \cite{2015MNRAS.451..849A} measured a halo mass of $\log_{10}(M_{\rm h} /h^{-1} M_\odot)=13.5^{+0.5}_{-1.5}$ for a similar RLAGN sample, albeit with a higher average redshift $z=0.5$. \cite{2004MNRAS.350.1485M} estimate a halo mass of $\log_{10}(M_{\rm h} /h^{-1} M_\odot)=13.8^{+0.2}_{-0.3}$ at $z = 0.11$ through a clustering analysis for a RLAGN sample also selected through FIRST. Although our measurement is at the high end, the results are statistically consistent.

Measuring a single `average' halo mass hides information about the more complex luminosity and redshift relationships that may exist, and the intrinsic halo mass distribution at fixed redshift and luminosity. For example, \cite{2004MNRAS.350.1485M} infer that there is a halo mass cut-off for RLAGN with an estimate of $\log_{10}(M_{\rm h} /h^{-1} M_\odot)\gtrsim13.5$ at $z \approx 0.1$. We can actually test the influence of massive clusters on our halo mass estimate by removing RLAGN that coincide with the positions of known clusters with $\log_{10}(M_{\rm h} /h^{-1} M_\odot)>14.0$. Using the redMaPPer catalogue \citep{2014ApJ...785..104R} we repeated our analysis after removing RLAGN that lie within 1-arcminutes of an optically-selected cluster, corresponding to 
15\% of the sample. Excluding these, the resulting bias drops to $b=2.0 \pm 0.8$ (at the same redshift) corresponding to an average halo mass of $\log_{10}(M_{\rm h} /h^{-1} M_\odot)=13.7^{+0.4}_{-1.0}$, comparable to the results of \cite{2004MNRAS.350.1485M} and \cite{2015MNRAS.451..849A} and indicating that RLAGN populate the massive end of the halo mass function.
 
Although we find broadly consistent results, a key difference between our sample and \cite{2004MNRAS.350.1485M} and \cite{2015MNRAS.451..849A} is a higher luminosity cut. The flux density limit of $S_{\rm 1.4\,GHz}>5$\,mJy selects galaxies with $L_{\rm 1.4GHz} \approx 10^{23}{\rm W\,Hz^{-1}}$ at $z=0.1$ and our RLAGN sample is classified via spectroscopy. \cite{2015MNRAS.451..849A} select with $S_{\rm 1.4\,GHz}>1$\,mJy and assume that sources at $z>0.2$ are RLAGN. \cite{2004MNRAS.350.1485M} also probe a lower luminosity range $10^{20} \lesssim (L_{\rm 1.4GHz}/{\rm W\,Hz^{-1}}) \lesssim 10^{24}$ and identify all the RLAGN as FRI, however, they found little dependency of RLAGN halo mass on luminosity. Unfortunately, the current sample is too small to probe any redshift or luminosity dependence in the lensing cross-power signal, although this will become possible with future large radio surveys with LOFAR, ASKAP, MeerKAT and eventually the SKA, provided reliable redshifts and classifications can be made. 

\section*{Acknowledgements}
CD is grateful to the Daphne Jackson Trust and the UK Science and Technology Facilities Council (STFC) who fund and support this work through a Daphne Jackson Research Fellowship. JEG acknowledges the support of the Royal Society through a University Research Fellowship. MJH acknowledges support from STFC [ST/M001008/1]. 
This research has made use of the University of Hertfordshire high-performance computing facility \url{http://uhhpc.herts.ac.uk}. This research made use of {\tt astropy}, a community-developed core Python package for astronomy \citep{AstropyCollaboration:2018ti}, and of {\it topcat} \citep{2005ASPC..347...29T}. The work is based on observations obtained with {\it Planck} \url{http://www.esa.int/Planck}, an ESA science mission with instruments and contributions directly funded by ESA Member States, NASA, and Canada.




\bibliographystyle{mnras}
\bibliography{rlagnlibrary} 

\bsp	
\label{lastpage}
\end{document}